\documentclass[onecolumn]{elsart}
\usepackage{amssymb}
\usepackage{amsfonts}
\usepackage{amsmath}
\usepackage{graphicx}
\usepackage{subfigure}
\usepackage{epsfig}
\usepackage{hyperref}
\usepackage{appendix}
\usepackage{cite}

\input{tcilatex}

\begin{document}

\begin{frontmatter}%

\title{Revisiting the local potential approximation of the exact
renormalization group equation}%

\author{C.\ Bervillier}\ead{claude.bervillier@lmpt.univ-tours.fr}%

\address{Laboratoire de Math\'{e}matiques et Physique Th\'{e}orique,\\ UMR 7350 (CNRS),\\
F\'ed\'eration Denis Poisson,\\
Universit\'{e} Fran\c{c}ois Rabelais,\\
Parc de Grandmont, 37200 Tours, France}%

\begin{abstract}
The conventional absence of field renormalization in the local potential
approximation (LPA) --implying a zero value of the critical exponent $\eta $%
-- is shown to be incompatible with \ the logic of the derivative expansion
of the exact renormalization group (RG) equation. We present a LPA with $%
\eta \neq 0$ that strictly does not make reference to any momentum
dependence. Emphasis\ is made on the perfect breaking of the
reparametrization invariance in that pure LPA (absence of any vestige of
invariance) which is compatible with the observation of a progressive smooth
restoration of that invariance\ on implementing the two first orders of the
derivative expansion whereas the conventional requirement ($\eta =0$ in the
LPA) precluded that observation.
\end{abstract}%


\end{frontmatter}

\section{Introduction}

The exact renormalization group (RG) equation \cite{440} (ERGE) --also
called non-perturbative or functional RG equation-- cannot be concretely
used without recourse to approximation (for modern reviews or introductory
lectures see, e.g., \cite{4595,4700,4823}). The best known approximation
framework for the ERGE is the derivative expansion \cite{212,4429,3836}. The
leading order of that expansion, $O\left( \partial ^{0}\right) $-order, also
named the local potential approximation (LPA) \cite{3480,4398,2085,2080},
completely discards any momentum dependence from the study. In principle the
LPA amounts to projecting the RG flow of the complete action $S\left[ \phi %
\right] $ (a functional of the field $\phi \left( x\right) $) onto the space
of simple functions $U\left( \phi \right) $ of a uniform field $\phi $ by
assuming that:%
\begin{equation}
S\left[ \phi \right] =\Omega _{D}\,U\left( \phi \right)  \label{1}
\end{equation}%
where $\Omega _{D}$ is the volume of the $D$-dimensional space.

Due to its simplicity and because it is thought that it qualitatively
involves most of the properties of the complete ERGE in the large distance
regime (e.g., stability properties and number of fixed points), the LPA is
currently utilized in many studies. Numerically, the LPA is considered as a
reasonable approximation because the estimations of the critical properties
would only be vitiated by the obligatory zero value of the critical exponent 
$\eta $ (characterizing the large distance behavior of the two-point
correlation function at the critical point) which, in many circumstances, is
actually a small parameter.

In the early studies, the condition $\eta =0$ in the LPA has been justified
as a consequence of the neglect of the detailed momentum dependence in the
RG\ \cite{3480} (the same kind of justification of $\eta =0$ may be found in 
\cite[p. 121]{440}). Though this argumentation by default is sometimes
reused \cite{2080,3491}, it is not very strong. It has been argued that, in
the LPA, \textquotedblleft \textsl{it is not possible to consistently
determine }$\eta $\textquotedblright , or $\eta $ \textquotedblleft \textsl{%
is set to zero as there is no mechanism to determine}\textquotedblright\ its
value\ \cite{6408} (see also \cite{7289}). The alert reader could express
some surprise and argue that the vanishing of $\eta $ in the LPA has been
clearly demonstrated a long time ago by Hasenfratz and Hasenfratz \cite{2085}
as often put forward in current studies (see, e.g, \cite{3553,3550,4595}).
Unfortunately, the arguments are not unassailable because they rely, at
least allusively (see section \ref{Invalidity}), on the following truncation
of $S\left[ \phi \right] $:%
\begin{equation}
S\left[ \phi \right] =\Omega _{D}\,U\left( \phi \right) +\frac{\bar{z}}{2}%
\int d^{D}x\left( \partial _{x}\phi \left( x\right) \right) ^{2}  \label{2}
\end{equation}%
in which the coefficient $\bar{z}$ of the kinetic term would be maintained
unaltered (equal to unity) along a RG flow of $U$. A condition which would
imply $\eta =0$ \cite{2085}. Pending to show that the argument is actually
artificial (see section \ref{Invalidity}), we may already notice that
truncation (\ref{2}) differs in nature from the pure LPA (\ref{1}) since it
refers partly to the $O\left( \partial ^{2}\right) $-order of the derivative
expansion. Consequently, assuming it was correct, this currently accepted
argument, basis of what is referred to in the following as the conventional
LPA, spoils the logic of the expansion based on a systematic projection of
the complete ERGE onto the space of actions successively truncated according
to the number of the derivatives of the field $\phi \left( x\right) $.
Normally the LPA should correspond to (\ref{1}) and not to (\ref{2}) so that
the supposedly proof of \cite{2085}, even true, would be inappropriate.
Hence, only remains the poor default argument. It is then legitimate to
wonder whether the condition $\eta =0$ is actually obligatory in the pure
LPA.

It is a matter of fact that the conventional value $\eta =0$ in the LPA
--not accidentally but associated with a systematic absence of field
renormalization-- raises some questions:

\begin{enumerate}
\item The absence of any field renormalization precludes in the LPA the
eventual setting up of a non-classical power law behavior of correlation
functions at criticality \cite{2727} other than that purely induced by a
diverging correlation length. According to the theory of critical phenomena,
two critical exponents are necessary to determine all the other critical
indices of a second order critical point. With the ERGE considered around a
Wilson-Fisher-like (W\thinspace F)\footnote{%
A non-trivial fixed point with one direction of infra-red instability
associated with the critical exponent $\nu $ of the correlation length.}
fixed point \cite{439}, these two exponents are $\eta $ and $\nu $. The two
exponents arise differently in the ERGE. The index $\nu $ (which
characterizes the divergence of the correlation length $\xi $ when the
temperature $T$ approaches its critical value $T_{c}$) occurs as a positive
eigenvalue of the RG equation linearized around a fixed point (the number of
such positive values determines the order of the transition). The role of
the index $\eta $ is more subtle. It is associated with the field
renormalization allowing for a non-classical power law behavior of the
correlation function at criticality. Usually one introduces $\eta $ in order
to reproduce the critical behavior of the correlation function at large
distances (momenta going to zero) and $T=T_{c}$; this manner of doing
tightly links the field renormalization to the momentum dependence and
suggests that no field renormalization is required when the momentum
dependence is neglected \cite{440,3480} implying $\eta \equiv 0$. However,
the fluctuation-dissipation theorem relates the correlation function to the
susceptibility. This allows another introduction of $\eta $ via the critical
exponent $\gamma =\nu \left( 2-\eta \right) $ characterizing the critical
behavior of the two-point correlation function at zero momenta and $%
T\rightarrow T_{c}$. No reference to an explicit momentum is required in
that case but the field should be renormalized nonetheless (to take this
eventual non-trivial power law with $\gamma \neq 2\nu $ into account). In
the LPA (conventional or not) $\nu $ takes on a non-classical value \cite%
{6137} when the fixed point is a non-trivial one. Then, there is a priori no
reason why $\eta =2-\gamma /\nu $ would take on\ a classical value at this
fixed point (a priori no reason for keeping the field unrenormalized).

\item \label{Invariance}Although broken by the derivative expansion, the
reparametrization invariance of the complete ERGE is expected to be
progressively restored as the order of the expansion grows. When it is
satisfied, this invariance specifies, in particular, that a change of
normalization of the field by a pure constant [like the parameter $\bar{z}$
in (\ref{2})] generates a line of equivalent fixed points characterized by a
unique set of critical exponents with the joint existence of a zero
eigenvalue mode in the solutions of the ERGE linearized around those fixed
points. The breaking of that invariance by the derivative expansion has been
concretely observed at next-to-leading order [$O\left( \partial ^{2}\right) $%
-order] \cite{212,3836,5744}; it is such that, for a given smooth cutoff
function, a line of fixed points is well generated by the change of
normalization of $\phi $ but those fixed points are not equivalent.
Nonetheless, in agreement with a remark of Bell and Wilson \cite{4420} in
such a situation, one observes the existence at $O\left( \partial
^{2}\right) $-order of a vestige of the invariance via an extremum\footnote{%
On varying the normalization of the field $\phi $ by a constant factor $z$.}
of $\eta $ \cite{212,3836,5744} accompanied by the presence of a zero mode.
This gives a preferred estimate for $\eta $ (one sometimes also refers to a
principle of minimal sensitivity\footnote{%
In the process of a partial restoration of the reparametrization invariance,
this principle is not always efficient because the search for a concomitant
zero mode is not systematic (see section \ref{PMS}).}, see, e.g, \cite%
{2267,5469,5805}). Because the conventional LPA offers no opportunities to
look at the state of the invariance\footnote{$\eta $ being arbitrarily fixed
to zero there is no line of fixed points to be observed.}, then no signs of
progressive restoration of the invariance may be observed by going from the $%
O\left( \partial ^{0}\right) $-order to the $O\left( \partial ^{2}\right) $%
-order of the derivative expansion. Thus, considered as the leading order of
that expansion, the conventional treatment creates confusion about the
convergence property of the derivative expansion. This is a pity because the
issue is of some importance.

\item \label{EtaDet}Having defined the RG-time $t=-\ln \left( \Lambda
/\Lambda _{0}\right) $ (with $\Lambda $ the running cutoff scale and $%
\Lambda _{0}$ an arbitrary fixed momentum scale) the critical exponent $\eta 
$ is actually defined in the RG as the limit of a function $\eta \left(
t\right) $ on approaching a given fixed point (when $t\rightarrow +\infty $%
). According to Wilson's prescriptions \cite{440}, the function $\eta \left(
t\right) $ is determined by keeping fixed the coefficient of one
\textquotedblleft particular\textquotedblright\ term in the action $S\left[
\phi ,t\right] $ with the initial condition $\eta \left( 0\right) =0$. It is
customary to keep constant the coefficient $\bar{z}$ of the kinetic term
because, due to a symmetry of the action linked to the reparametrization
invariance, the flow of such a term is non-essential (redundant) so that it
may be constrained without altering the model integrity. When the kinetic
term is not part of the approximation, as in the pure LPA, the redundancy
still exists at least formally and it seems logical to wonder what the state
of the reparametrization process is in the pure LPA. To this end one should
introduce a function $\eta \left( t\right) $ that would maintain fixed a
\textquotedblleft particular\textquotedblright\ monomial of $U\left( \phi
\right) $. The line of fixed points mentioned in point \ref{Invariance}
would presumably be generated and this would give back the status of genuine
leading order [$O\left( \partial ^{0}\right) $-order] of the derivative
expansion to the LPA (see section \ref{RevisedLPA}).
\end{enumerate}

In section \ref{RevisedLPA} we present and discuss a version of the LPA with 
$\eta \neq 0$. We show that it satisfies all the required conditions for
being a genuine $O\left( \partial ^{0}\right) $-order of the derivative
expansion. In particular we study, in section \ref{FixedPoints}, the
structure of the fixed points for any value of the dimension $D$ and show
explicitly, for the first time in the LPA, how the reparametrization
invariance is broken. We also introduce, in section \ref{LT}, a Legendre
transformation of the potential adapted to the case studied ($\eta \neq 0$).
This allows us to utilize easy quasi-analytic methods (section \ref{TSM}) of
integration of an ordinary differential equation (ODE) well adapted to the
obtention of the eigenvalues of the RG flow linearized around a fixed point.
It is then shown that the principle of minimal sensitivity (PMS) does not
necessarily indicate preferred values of the critical exponents (section \ref%
{PMS}). In section \ref{VersusPseudoLPA}, we first show that the
conventional argument which is generally put forward to justify $\eta =0$ in
the LPA, is actually artificial (section \ref{Invalidity}). We then discuss
briefly, according to the RG rules, how the reparametrization invariance
could be studied\ with the partial truncation (\ref{2}) (section \ref%
{PseudoLPA}). In appendix \ref{Movable}, we illustrate some reason why the
quasi-analytic methods of integration of ODE of section \ref{TSM} do not
work in the case of the potential of the action whereas they work after a
Legendre transformation is performed. We conclude in section \ref{Conclusion}%
.

\section{The revised LPA\label{RevisedLPA}}

Let us consider the RG flow equation of the Polchinski ERGE extended to
include the parameter $\eta \neq 0$ in the LPA \cite{3491}, it reads%
\begin{equation}
\dot{U}=U^{\prime \prime }-U^{\prime 2}-\frac{D-2+\eta \left( t\right) }{2}%
\phi U^{\prime }+D\,U  \label{LPA0}
\end{equation}%
where $U\left( \phi ,t\right) $ stands for a simple function of $\phi $ and $%
t$, $\dot{U}\equiv \left. \partial U/\partial t\right\vert _{\phi }$, $%
U^{\prime }=\left. \partial U/\partial \phi \right\vert _{t}$, $U^{\prime
\prime }=\left. \partial ^{2}U/\partial \phi ^{2}\right\vert _{t}$, $D$ is
the spatial dimension, and $\eta \left( t\right) $ the field renormalization
parameter which, at a fixed point, takes on the value $\eta ^{\ast }$. In
principle and with the complete ERGE, $\eta ^{\ast }$ should coincide with
the critical index $\eta $.

The flow equation (\ref{LPA0}) has already been studied by Kubyshin et al 
\cite{4222,4988,5255} for the derivative $f\left( \phi ,t\right) =U^{\prime
} $. But they have considered $\eta \left( t\right) \neq 0$ in the LPA for
technical reasons exclusively \cite{4222}. Hence, in accordance with the
conventional LPA, they have left $\eta \left( t\right) $ undefined and
focused their interest on $\eta ^{\ast }$ considered as an arbitrarily
adjustable parameter while emphasizing that physically $\eta ^{\ast }$
should be zero at this order of the derivative expansion.

With a view to study the fixed point equation ($\dot{U}=0$) for any $D$ at
one time, we perform the following change of normalization of\ $\phi $:%
\begin{equation}
\phi \rightarrow \frac{\phi }{\sqrt{D}}  \label{ChangePhi}
\end{equation}%
then Eq. (\ref{LPA0}) transforms into \cite{2080}:%
\begin{eqnarray}
\dot{U} &=&D\,\left[ U^{\prime \prime }-U^{\prime 2}-\mu \left( t\right)
\phi U^{\prime }+U\right]  \label{LPA} \\
\mu \left( t\right) &=&\frac{D-2+\eta \left( t\right) }{2D}  \label{mu}
\end{eqnarray}%
For a given $D$, $\mu \left( t\right) $ plays the role of $\eta \left(
t\right) $ and for any $D$, the fixed point equation involves only one
parameter instead of two in the preceding case of (\ref{LPA0}).

Considering exclusively the issue of finding a non-singular solution $%
U^{\ast }\left( \phi \right) $ to the fixed point equation corresponding to (%
\ref{LPA}) [or (\ref{LPA0})], gives no possibility for determining a value
of $\mu ^{\ast }=\left( D-2+\eta ^{\ast }\right) /\left( 2D\right) $. Then $%
\mu ^{\ast }$ may rightly be considered as an extra parameter. Indeed $%
U^{\ast }\left( \phi \right) $ is a solution of the following two-point
boundary value problem of a second order non-linear ODE:%
\begin{eqnarray}
U^{\ast \prime \prime }-U^{\ast \prime 2}-\mu ^{\ast }\,\phi U^{\ast \prime
}+U^{\ast } &=&0  \label{FP} \\
U^{\ast \prime }\left( 0\right) &=&0  \label{cond1} \\
U^{\ast \prime \prime }\left( \infty \right) &=&-\frac{1}{2}+\mu ^{\ast }
\label{cond2}
\end{eqnarray}%
Now the whole of the two integration constants are fixed by the property of
parity (\ref{cond1}) and by the adjustment of $U^{\ast }\left( 0\right)
=k^{\ast }$ so\ as to get a non-singular $U^{\ast }\left( \phi \right) $ in
the whole range $\phi \in \left] -\infty ,+\infty \right[ $ as prescribed by
condition (\ref{cond2}); then there is no room for determining $\mu ^{\ast }$
($\eta ^{\ast }$ at fixed $D$) without a supplementary condition.

In the conventional LPA, the supplementary condition is merely $\eta \left(
t\right) \equiv 0$ which would be obtained (see section \ref{Invalidity}
however) by an explicit reference to a larger space of truncation functions
[see Eq. (\ref{2})]. However, even correct, this procedure would not be
justified because the RG theory gives precise rules to determine both the
function $\eta \left( t\right) $ and its fixed point value $\eta ^{\ast }$.
As recalled in point \ref{EtaDet} of the introduction, the function $\eta
\left( t\right) $ is determined by keeping one particular term of the action
fixed along the RG flows \cite{440}; then $\eta ^{\ast }$ is the value
reached by $\eta \left( t\right) $ in approaching a given fixed point. This
procedure is a direct consequence of the reparametrization invariance of the
complete action which induces the redundancy of the flow of one term of the
action. In absence of any kinetic term, as in the pure LPA, it is logical,
and coherent with the RG theory, to define $\eta \left( t\right) $ by
keeping constant the coefficient of the quadratic term $U^{\prime \prime
}\left( \phi =0,t\right) .$

Let us examine, on a general ground, the approach of the W\thinspace F fixed
point $U_{WF}^{\ast }\left( \phi \right) $ with the flow equation (\ref{LPA}%
) starting at $t=0$ with the following simple potential:%
\begin{equation}
U\left( \phi ,0\right) =k_{0}+\frac{z}{2}\phi ^{2}+\frac{g_{0}}{4!}\phi ^{4}+%
\frac{u_{0}}{6!}\phi ^{6}  \label{VdePhi0}
\end{equation}%
where the coefficient of the quadratic term has been intentionally noted $z/2
$.

Because $U_{WF}^{\ast }\left( \phi \right) $ has only one relevant
direction, to make the flow approaching $U_{WF}^{\ast }\left( \phi \right) $
only one coefficient of (\ref{VdePhi0}), say $k_{0}$, must be fine-tuned (in
terms of the other three coefficients) \cite{4627}. This adjustment is
necessary to place the initial potential on the critical surface (within the
domain of attraction) of $U_{WF}^{\ast }\left( \phi \right) $ \cite{4627}.
In order to follow a RG flow, the function $\eta \left( t\right) $ must be
defined. We do it such that $U^{\prime \prime }\left( 0,t\right) =z$ all
along the RG flow, with the initial condition $\eta \left( 0\right) =0$ and $%
z$ a constant independent of $t$. At the fixed point (reached at infinite
RG-time provided the initial potential lies in the domain of attraction of
the fixed point), $\eta \left( \infty \right) $ takes on the value of $\eta
^{\ast }\left( z\right) $ and this defines a line of fixed points
(parametrized by $z$). If it was satisfied, the reparametrization invariance
would imply that $\eta ^{\ast }\left( z\right) $ be independent of $z$ and
equal to $\eta $. Of course, in the pure LPA one rather expects to observe
the breaking of that marvelous property and the\ true question is: to which
extent is that invariance broken in the LPA?

To look at this question, suffices to express the variation of $\eta ^{\ast
} $ in terms of $U^{\ast \prime \prime }\left( 0\right) $. This is precisely
what Kubyshin et al have done in \cite{4222,5255}: they studied Eq. (\ref%
{LPA0}) for $D=2$ and $3$ (for $D=3$ in \cite{4988}). The purpose of
Kubyshin et al was not the status of the reparametrization invariance in the
LPA however. In fact, having considered the flow equation for the derivative 
$f\left( \phi ,t\right) =U^{\prime }$, the connection parameter of their
fixed point equation was not $U^{\ast }\left( 0\right) =k^{\ast }$ but
instead $U^{\ast \prime \prime }\left( 0\right) $ that they have noted $%
\gamma $. Then they have naturally drawn the variation of $\gamma $ on
changing the value of $\eta ^{\ast }$ (or the reverse) without relating this
variation to the reparametrization process. It is however clear that, with
our prescription of keeping $U^{\prime \prime }\left( 0\right) =z$ fixed
along a RG flow, the fixed point is reached with $U^{\ast \prime \prime
}\left( 0\right) \equiv z$ where obviously $z$ may be considered as the
normalization of the field. Consequently, the evolutions of $\eta ^{\ast
}\left( \gamma \right) $ drawn by Kubyshin et al are nothing but
illustrations of the breaking of the reparametrization invariance in the
LPA. Let us redo the study of Kubyshin et al using our own conventions.

\subsection{Lines of Fixed points\label{FixedPoints}}

\begin{figure}[tbp]
\begin{center}
\includegraphics*[width=12cm]{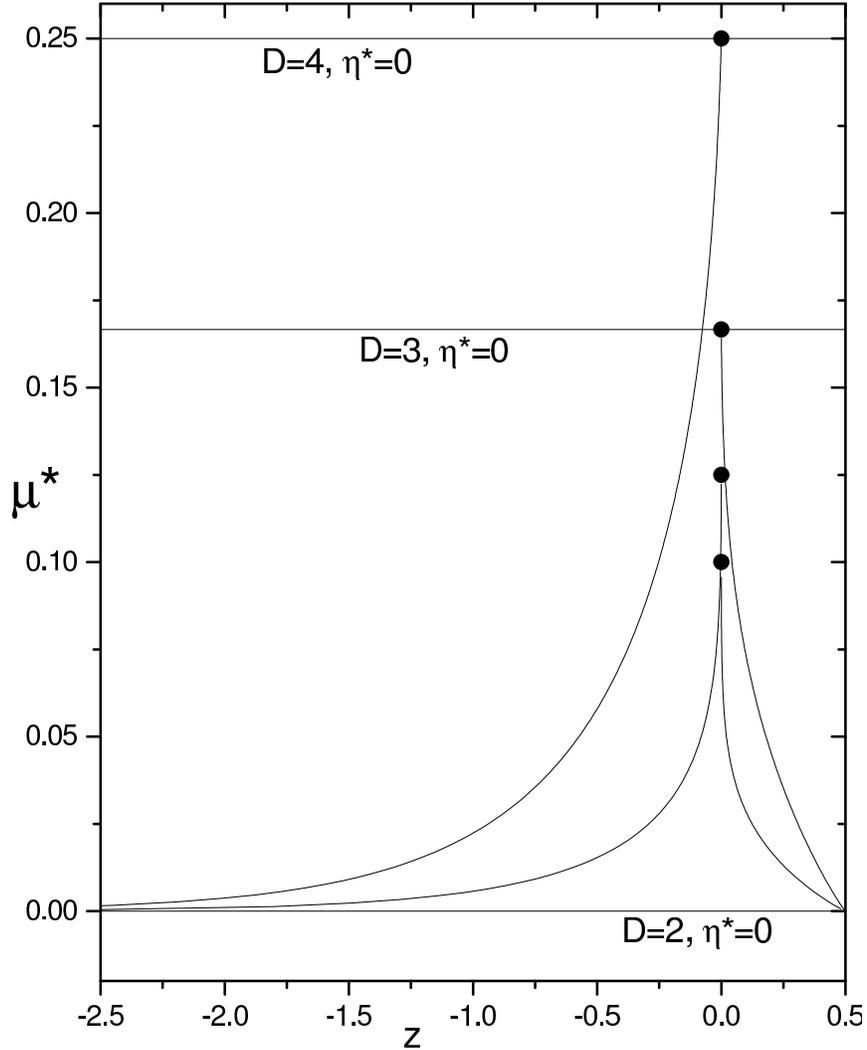}
\end{center}
\caption{Lines of fixed points $\protect\mu _{1}^{\ast }\left( z \right) $ to $\protect\mu _{4}^{\ast }\left( z \right) $ obtained as
regular solutions of Eqs. (\protect\ref{FP}-\protect\ref{cond2}). The full
circles represent the thresholds of instability of the Gaussian fixed point.
The horizontal lines indicate three values of $D$ where such instabilities
occur when $\protect\eta^* =0.$ See text for more details.}
\label{figPF1}
\end{figure}

Using a standard numerical shooting method, we have looked for regular
solutions (the values of $k^{\ast }=U\left( 0\right) $) of the two-point
boundary value problem (\ref{FP}-\ref{cond2}). Clearly, those solutions are
parametrized by $k^{\ast }\left( \mu ^{\ast }\right) $. Since the
coefficient of the quadratic term $U^{\ast \prime \prime }\left( 0\right)
\equiv z$ is linked to $k^{\ast }\left( \mu ^{\ast }\right) $ via the
differential equation as $z=-k^{\ast }$, one easily gets functions $\mu
_{n}^{\ast }\left( z\right) $ corresponding to the functions\footnote{\label%
{gamma}Due to (\ref{ChangePhi}) the value of $U^{\ast \prime \prime }\left(
0\right) =\gamma $ in the study of Kubyshin et al is related to our $z$ as $%
\gamma =D\,z$.} $\eta _{n}\left( \gamma \right) $ of Kubyshin et al \cite%
{5255}. The four first solutions are displayed in fig (\ref{figPF1}) as
continuous curved lines; this figure involves simultaneously the two graphs
of fig. (1) of Kubyshin et al \cite{5255} and displays the same features.
Let us discuss them.

Each curved line drawn on fig. (\ref{figPF1}) corresponds to a line of fixed
points of a particular nature. It appears as bifurcating from the Gaussian
fixed point (full circles) on varying $\mu ^{\ast }$ each time $\mu ^{\ast }$
falls below the thresholds $\mu _{n}^{\ast }=1/\left[ 2\left( 1+n\right) %
\right] $, $n=1,2,\ldots $ [horizontal lines on fig. (\ref{figPF1}),
corresponding to the usual dimensional thresholds $D_{n}=2+2/n$ (for $\eta
^{\ast }=0$)]. In particular fig. (\ref{figPF1}) shows the well-known fact
that the Gaussian fixed point is stable for $D>4$ and $\eta ^{\ast }=0$ ($%
\mu ^{\ast }>1/4)$ where there is no regular solution to Eq. (\ref{FP}). A
new fixed point bifurcates each time the Gaussian fixed point acquires a new
direction of instability; then the fixed points belonging to a line $\mu
_{n}^{\ast }\left( z\right) $ have $n$ directions of instability each. Fig. (%
\ref{figPF1}) shows the four first lines of fixed points $\mu _{1}^{\ast
}\left( z\right) $ to $\mu _{4}^{\ast }\left( z\right) $) having
respectively one, two, three and four directions of instability. All the
lines accumulate at the horizontal line $\mu ^{\ast }=0$ as $n\rightarrow
+\infty $ along which $z$ reaches $-\infty $ or stop at $1/2$ in agreement
with the analytical solution found by Kubyshin et al \cite{5255} for $\mu
^{\ast }=0$.

Let us focus our attention on the line $\mu _{1}^{\ast }\left( z\right) $
which is a line of W\thinspace F fixed points. Because, for a given $D$, $%
\eta ^{\ast }$ varies along the line, it is obvious that the
reparametrization invariance is broken. Of course, this was expected in the
LPA but surprisingly had never been explicitly emphasized before the present
study. Because the line is smooth and monotonous, there is no vestige of the
invariance, the breaking is perfect except at the limiting value $\mu ^{\ast
}=0$ where, for $D=3$, $\eta ^{\ast }$ takes on the values $-1$. Notice that
nothing particular distinguishes the value $\eta ^{\ast }=0$ from the other
values\footnote{%
The usual W-F fixed point of the conventional LPA with $D=3$ lies at the
intersection of $\mu _{1}^{\ast }\left( z\right) $ and the horizontal line $%
D=3$, $\eta ^{\ast }=0$ with $z\simeq -0.07620$ (i.e. $\gamma \simeq -0.2286$%
).} except at the limiting cases $\mu ^{\ast }=0$, $D=2$ and $\mu ^{\ast
}=1/4$, $D=4$.

One observes also that $z<0$ on the whole line of fixed points $\mu
_{1}^{\ast }\left( z\right) $ (except the Gaussian fixed point); this means
that the basin of attraction of a non-trivial W\thinspace F fixed point
implies the condition $U^{\prime \prime }(0,0)<0$ on the initial potential $%
U\left( \phi ,0\right) $ (otherwise the RG flow goes away from the critical
surface towards the trivial high-temperature fixed point). Notice that the
perfect breaking of the reparametrization invariance does not completely
spoil the universal character of the critical behavior since the infinite
number of initial potentials with a given $U^{\prime \prime }(0,0)<0$ lying
on the critical surface are characterized by the same critical behavior
governed by a unique value of $\eta ^{\ast }$ and, subsidiarily, of the
other critical exponents.

From the line $\mu _{1}^{\ast }\left( z\right) $, it is also interesting to
notice that non-trivial fixed points may be formally considered as existing
for $D=4$ provided that $\eta ^{\ast }$ be strictly negative. Usually such
fixed points are rejected but in the present study there is no reason to
reject them\ a priori since it is a consequence of the breaking of the
reparametrization invariance to generate also negative values of $\eta
^{\ast }$. It would be puzzling however if a vestige of that invariance led
us to choose such a negative value of $\eta ^{\ast }$. Fortunately we
observe no sign of such a preferred value of $\eta ^{\ast }$ along the line $%
\mu _{1}^{\ast }\left( z\right) $ except the limit case $\mu ^{\ast }=0$.

At this level we conclude that the LPA does not allow one to determine any
estimate of $\eta ^{\ast }$ but only ranges of possible values. For example,
if one excludes negative values of $\eta ^{\ast },$ then for $D=3$ this
range would be $\left[ 0,1/2\right[ $\ for the only line $\mu _{1}^{\ast }$,
the other lines being excluded. From this example taken alone, one could be
inclined to conclude that the conventional LPA, by imposing $\eta ^{\ast }=0$%
, would be merely a reasonable choice since one knows that $\eta ^{\ast }$
is most often small. However, fig. (\ref{figPF1}) shows that from $D<3$ down
to $D=2$, emerges a rich structure of various fixed points with possible
different positive and growing values of $\eta ^{\ast }$ for which the
conventional LPA would impose, increasingly poorly as $D$ decreases, the
same zero value (one knows that at $D=2$, $\eta =1/4$ what is not small).

Notice that, for $D=3,$ we have excluded the limit case $\eta ^{\ast }=1/2$ 
\footnote{%
Similarly, $\eta ^{\ast }=0$ on the line $\mu_2^*$.} from the range of
possible values of $\eta ^{\ast }$ though it corresponds to the only point
on $\mu _{1}^{\ast }\left( z\right) $ where a zero eigenvalue exists (see
section \ref{PMS}). Indeed, since \textrm{d}$\eta ^{\ast }/\mathrm{d}z\neq 0$
at this point, this zero mode is not a vestige of the reparametrization
invariance; instead it indicates that the nature of the Gaussian fixed point
is going to change by losing one direction of instability. Hence, if a
direct derivative of the fixed point equation with respect to $z$ shows that
an extremum of $\eta^*$ implies the appearance of a zero eigenvalue, the
reverse is not true.

In order to better illustrate the role of the zero mode in the process of
restoration of the reparametrization invariance, let us look at the critical
exponents in the LPA and at their variations on changing the normalization
of the field. To this end, we perform a Legendre transformation of the
potential which will allow us to make use of user-friendly quasi-analytic
methods of \textquotedblleft integration\textquotedblright\ of ODE.

\subsection{ Eigenvalues: Taylor series, Legendre transformation, principle
of minimal sensitivity}

\subsubsection{Taylor series methods\label{TSM}}

The interest of using some quasi-analytic methods to solve the RG flow
equation in the LPA is the extremely easy access that they offer to estimate:

\begin{enumerate}
\item the fixed point value $k^{\ast }$ of the connection parameter

\item a set of critical (and subcritical) exponents at one time.
\end{enumerate}

On the contrary, the purely numerical shooting method necessitates a skillful
adjustment of an initial guess of the final value of $k^{\ast }$ or,
independently, of each critical exponent sought.

The use of quasi-analytic methods based on Taylor series, in solving a
two-point boundary value problem like (\ref{FP}-\ref{cond2}), has been
recently reviewed and illustrated in \cite{6889}. Among such methods is an
extremely simple procedure \cite{3478,3553,4004} (named the simplistic
method in \cite{6889}) that merely consists of imposing the vanishing of the
last term $a_{M}\left( k\right) $ of the Maclaurin series of the truncated
solution: 
\begin{equation}
U_{M}\left( \phi \right) =k+\sum_{i=1}^{M}a_{i}\left( k\right) \phi ^{2i}
\label{Taylor}
\end{equation}%
the coefficients $a_{i}\left( k\right) $ being determined such that the EDO
considered be satisfied order by order in powers of $\phi ^{2}$. The
auxiliary condition $a_{M}\left( k\right) =0$ gives a condition from which
one tries to extract an estimate of the connection parameter $k^{\ast
}=U^{\ast }\left( 0\right) $ corresponding to the only regular solution of (%
\ref{FP}-\ref{cond2}). Of course, because it is too simple, this simplistic
method is not always (most often never) efficient. Firstly the finite
character of the radius of convergence of the series limits the accuracy of
the method \cite{3553,4004}. Secondly the method may simply not work at all
(in the sense that even a rough estimate of $k^{\ast }$ may not be
approachable). Indeed, in trying to solve (\ref{FP}-\ref{cond2}), the issue
we are faced with amounts to pushing a movable singularity to infinity. The
efficiency of the simplistic method then depends on whether or not that
singularity lies within the circle of convergence of the Maclaurin series or
not (see appendix \ref{Movable}).

A variant of the simplistic method, referred to below as the Taylor method,
is frequently used which is based on a Taylor expansion around the minimum
of the potential, as proposed in \cite{4192,3642} (see also \cite{3553,4004}%
). The solution of the ODE is thus expressed as:

\begin{eqnarray*}
U_{M}\left( \phi \right) &=&k_{0}+\sum_{i=2}^{M}b_{i}\left(
k_{0},x_{0}\right) \left( x_{0}-x\right) ^{i} \\
x &=&\phi ^{2} \\
x_{0} &=&\phi _{0}^{2}
\end{eqnarray*}%
where $\phi _{0}$ is the expansion point chosen to coincide with the minimum
of the potential\footnote{%
This choice is not obligatory. One could have fixed $x_{0}$ arbitrarily and
the two unknowns would have been $k_{0}$ and $b_{1}$. This would offer the
possibility of improving the apparent convergence of the Taylor method by
varying $x_{0}$, see \cite{4004}.} since $b_{1}\left( k_{0},x_{0}\right) =0$%
, and $k_{0}=U_{M}\left( \phi _{0}\right) $, whereas the original connection
parameter is, by definition, given by:%
\begin{equation*}
k=k_{0}+\sum_{i=2}^{M}b_{i}\left( k_{0},x_{0}\right) x_{0}^{i}
\end{equation*}%
There are two unknowns ($k_{0}$ and $x_{0}$) to be determined. The Taylor
method consists of imposing the vanishing of the two last terms of the
Taylor series to get two auxiliary conditions on $k_{0}$ and $x_{0}$. This
method may improve considerably the simplistic method (it has provided
excellent estimates of the critical exponents in the LPA \cite{6137}). The
reason is due to the fact that, other things being equal compared to the
simplistic method, one starts closer to the movable singularity. But the
Taylor method requires that the expansion point (the minimum of the
potential) lies within the circle of convergence of the Maclaurin series
(otherwise one could not get a reliable estimate of $k^{\ast }$ by summing
the series back to the origin). Also, the accuracy of the method is
naturally limited by the finite range of convergence of the Taylor expansion.

It is a matter of fact that, in the conventional LPA with $\eta \left(
t\right) \equiv 0$, the two quasi-analytic methods\footnote{%
The methods are not completely analytic because the determination of the
solution of the auxiliary condition is finally numerically performed.}
presented just above do not work when they are applied to the Polchinski RG
flow equation of $U$ but they work if one first performs a Legendre
transformation ($\left\{ U,\phi \right\} \rightarrow \left\{ V,\varphi
\right\} $) as that defined in \cite{5911}.

With a view to make use of these user-friendly quasi-analytic methods --that
allow anyone to easily verify the content of the present paper--, let us
introduce a Legendre transformation appropriate to the case $\eta \neq 0$.

\subsubsection{Legendre transformation for $\protect\eta \neq 0\label{LT}$}

To begin with, we consider the Legendre transformation originally introduced
for $\eta \left( t\right) \equiv 0$ in \cite{5911} and we apply it to the
flow equation of $V$ extended to include $\eta \neq 0$, namely:%
\begin{equation}
\dot{V}=\frac{V^{\prime \prime }}{1+V^{\prime \prime }}-\frac{D-2+\eta
\left( t\right) }{2}\varphi V^{\prime }+D\,V  \label{LPAMorris0}
\end{equation}

According to \cite{5911}, the Legendre transformation reads 
\begin{eqnarray}
U\left( \phi ,t\right) &=&V\left( \varphi ,t\right) +\frac{1}{2}\left( \phi
-\varphi \right) ^{2}  \label{TL1} \\
\varphi &=&\phi -U^{\prime }\left( \phi ,t\right)  \label{TL2} \\
\dot{U} &=&\dot{V}  \label{TL3}
\end{eqnarray}%
from which we have:%
\begin{eqnarray*}
U^{\prime } &=&V^{\prime } \\
U^{\prime \prime } &=&\frac{V^{\prime \prime }}{1+V^{\prime \prime }}
\end{eqnarray*}

Thus applied to (\ref{LPAMorris0}) we get the following flow equation for $U$%
: 
\begin{eqnarray}
\dot{U} &=&U^{\prime \prime }-\varpi \left( t\right) U^{\prime 2}-\frac{%
D-2+\eta \left( t\right) }{2}\phi U^{\prime }+D\,U  \label{LPA0bis} \\
\varpi \left( t\right)  &=&1-\frac{\eta \left( t\right) }{2}
\label{omgeaBar}
\end{eqnarray}%
which differs from the usual Polchinski equation (\ref{LPA0}) when $\varpi
\left( t\right) \neq 1$. The appearance of this coefficient may be seen as
the consequence of a non-linear introduction\footnote{%
As done originally in the historic first version \cite{440}.} of $\eta
\left( t\right) $ in the ERGE \cite{7849} instead of the linear introduction
of \cite{3491} that corresponds to (\ref{LPA0}). Though, near a fixed point,
the coefficient $\varpi ^{\ast }=1-\eta ^{\ast }/2$ may be removed from (\ref%
{LPA0bis}) through the change $U^{\ast }\rightarrow U^{\ast }/\varpi ^{\ast }
$ to get the same equation as (\ref{LPA0}), we have numerically studied%
\footnote{%
We could have considered instead a modified version of Eq. (\ref{LPAMorris0}%
) corresponding to applying the Legendre transformation (\ref{TL1}-\ref{TL3}%
) on (\ref{LPA0}), but this would have made the quasi-analytic methods
heavier and thus less attractive.} Eq. (\ref{LPA0bis}) explicitly for $D=3$
(using a shooting method). We have, this way, verified explicitly (in the
case of W\thinspace F fixed points) both that we get the same kind of line
of fixed points as previously (monotonous function $\eta ^{\ast }\left(
z\right) $) and that the simplistic and Taylor methods applied to (\ref%
{LPAMorris0}) work well also for $\eta \neq 0$ [at least for the values of $z
$ shown in fig (\ref{figNuDez})].

\subsubsection{Eigenvalues\label{PMS}}

Let us focus our interest on the eigenvalue problem corresponding to
Eq. (\ref{LPAMorris0}) linearized around a fixed point $V^{\ast }$
[solutions of (\ref{LPAMorris0}) such as $\dot{V}^{\ast }=0$]. We get the
following second order linear ODE (once $V^*$ is known):%
\begin{equation}
-\frac{v^{\prime \prime }}{\left( 1+V^{\ast \prime \prime }\right) ^{2}}-%
\frac{D-2+\eta ^{\ast }}{2}\varphi v^{\prime }+\left( D-\lambda \right) \,v=0
\label{VPLitim}
\end{equation}%
where $v\left( \phi \right) $ is the eigenfunction and $\lambda $ the
eigenvalue parameter.

For a given set of initial conditions, such as $v\left( 0\right) =1$, $%
v^{\prime }\left( 0\right) =0$ in the even case, one expects to obtain an
infinite set of discrete couples $\left\{ v_{n}\left( \phi \right) ,\lambda
_{n}\right\} $ ordered according to the magnitude of $\lambda _{n}$. The
number of positive values depends on the fixed point considered. Except the
trivial eigenvalue $\lambda _{0}=D$, the W\thinspace F fixed point is
characterized by the existence of only one positive value $\lambda _{1}$
corresponding to the critical exponent $\nu =1/\lambda _{1}$, the next
eigenvalue $\lambda _{2}$ is negative and corresponds to the leading
subcritical exponent $\Delta _{1}=\omega _{1}\,\nu $ characterizing the
leading correction-to-scaling with $\omega _{1}=-\lambda _{2}$. These two
exponents have been estimated, for $D=3$, with a very high accuracy in the conventional
LPA (with $\eta \equiv 0$) to get \cite{6137,6188,6319}:%
\begin{eqnarray*}
\nu  &=&0.6495617738806480176\cdots  \\
\omega _{1} &=&0.6557459391933387407\cdots 
\end{eqnarray*}%
Of course, due to the Legendre transformation, the same set of critical
exponents is obtained in both cases of the flow equations of $V$ and $U$ 
\cite{6137}. It is clear that this is also the case in the present study
with $\eta \neq 0$, provided the methods used converge.

\begin{figure}[tbp]
\begin{center}
\includegraphics*[width=10cm]{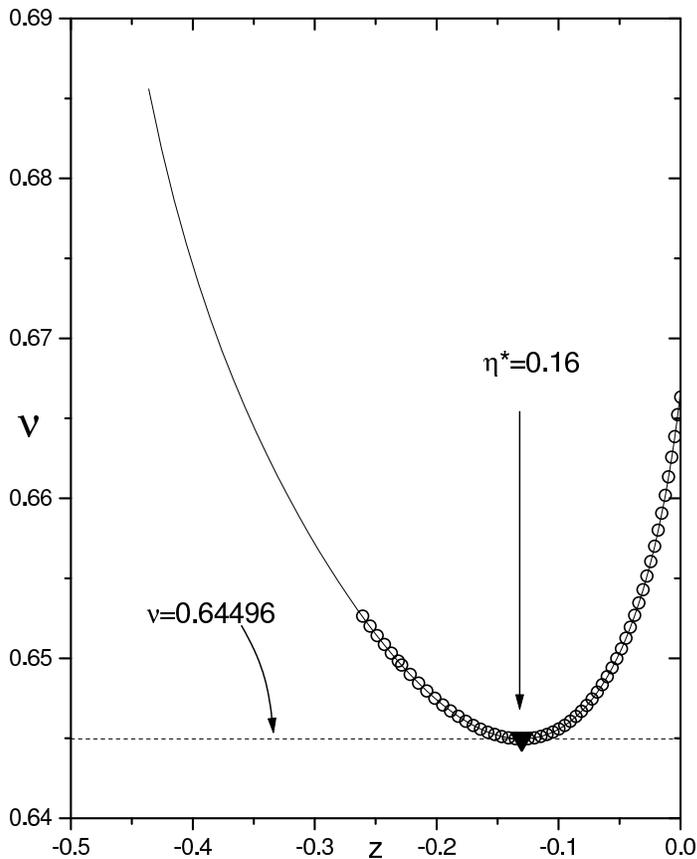}
\end{center}
\caption{Evolution, for $D=3$, of the critical exponent $\protect\nu $ as function of
the field-normalization $z$. The full line corresponds to calculations done
using the Taylor method, open circles correspond to results obtained with
the simplistic method. The point located at $z=0$ corresponds to the
Gaussian fixed point with $\protect\nu =\frac{2}{3}$.}
\label{figNuDez}
\end{figure}

\begin{figure}[tbp]
\begin{center}
\includegraphics*[width=10cm]{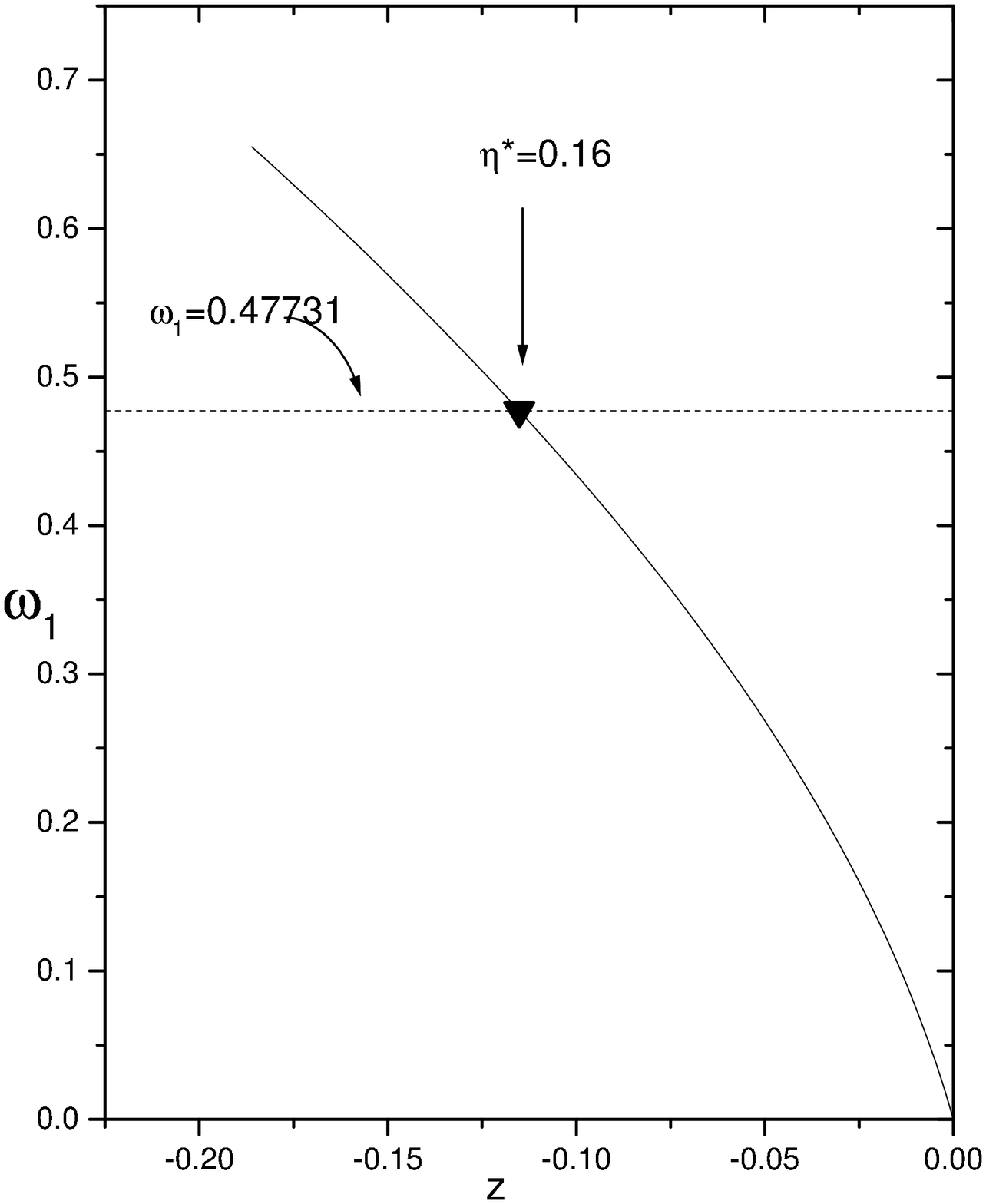}
\end{center}
\caption{Evolution, for $D=3$, of the subcritical exponent $\protect\omega _{1}$ as
function of the field-normalization $z$ (full line) obtained using the
Taylor method. The point located at $z=0,$ $\protect\omega _{1}=0$ is the
only possibility of having a zero mode; it is located far from the value
corresponding to $\protect\eta ^{\ast }=0.16$ for which $\protect\nu \left(
z\right) $ undergoes a minimum [see fig. (\protect\ref{figNuDez})].}
\label{figOmegaDez}
\end{figure}

We have determined the evolution in terms of $z=U^{\prime \prime }\left(
0\right) =V^{\prime \prime }\left( 0\right) /(1+V^{\prime \prime }\left(
0\right) )$ of the two first exponents $\nu $ and $\omega _{1}$ using both
the simplistic and Taylor methods and obtained the curves shown in figures (%
\ref{figNuDez}, \ref{figOmegaDez}).

Fig. (\ref{figNuDez}) shows that $\nu \left( z\right) $ undergoes a minimum
at $\nu _{\min }\simeq 0.64496$ corresponding to $\eta ^{\ast }\simeq 0.16$
whereas, at this point we get $\omega _{1}\simeq 0.47731$ [see fig. (\ref{figOmegaDez})]. According to a
principle of minimal sensitivity (PMS) \cite{2267} --sometimes used in
calculations at higher orders of the derivative expansion of the ERGE (see,
e.g, \cite{5469,5805})-- one could be inclined to propose those values as
being the preferred estimates of the critical exponents in the LPA for $D=3$. However,
one may observe that those values are not designated as the consequence of a
vestige of the reparametrization invariance which is the only reason that
fundamentally led us to vary $z$. Indeed no zero eigenvalue is obtained at
this point as shown in table \ref{Table1}. Fig. (\ref{figOmegaDez}) clearly
shows that the only point where a zero mode occurs corresponds to the
Gaussian fixed point which is losing one direction of instability ($\lambda
_{2}=0$, $\lambda _{1}=3/2$, $\eta ^{\ast }=1/2$, for $D=3$; or $\lambda
_{2}=0$, $\lambda _{1}=2$, $\eta ^{\ast }=0$, for $D=4$) but at this point $%
d\eta ^{\ast }/dz$ does not vanish [see fig. (\ref{figPF1})].

Notice that this is only a confirmation of the absence of any extrema in the
function $\eta ^{\ast }\left( z\right) $. Indeed, if one performs a
derivation with respect to $z$ of the fixed point equation corresponding to (%
\ref{LPAMorris0}), assuming \textrm{d}$\eta ^{\ast }\left( z\right) /\mathrm{%
d}z=0$, then one gets (\ref{VPLitim}) with $\lambda =0$. The reverse is not
true however: the presence of a zero mode may reveal instead the change of
the stability properties of the fixed point.

\begin{table}[tbp] \centering%
\begin{tabular}{l|l}
\hline\hline
Simplistic ($M=20$) & Taylor ($M=10$) \\ \hline
\multicolumn{1}{c|}{1.55050} & \multicolumn{1}{|c}{1.55049} \\ 
\multicolumn{1}{c|}{-0.47729} & \multicolumn{1}{|c}{-0.47731} \\ 
\multicolumn{1}{c|}{-2.7753} & \multicolumn{1}{|c}{-2.7773} \\ 
\multicolumn{1}{c|}{-5.243} & \multicolumn{1}{|c}{-5.231} \\ 
\multicolumn{1}{c|}{-7.91} & \multicolumn{1}{|c}{-8.49} \\ 
\multicolumn{1}{c|}{-10.4} & \multicolumn{1}{|c}{-13.4} \\ \hline
\end{tabular}%
\caption{Comparison of the six first eigenvalues of Eq. (\ref{VPLitim}) obtained for
$\eta^*=0.16$, $D=3$ and with the two quasi-analytic methods considered in the
study. The first line of numbers corresponds to $\lambda_1$  (1/$\nu$), the second to $\lambda_2$  ($-\omega_1$), etc. No zero eigenvalue is present.}%
\label{Table1}%
\end{table}%

We may thus conclude that, because it has no link with the reparametrization
invariance, the observed minimum of $\nu $ occurs accidentally and that the
PMS cannot be utilized in the circumstances as a tool to determine a
preferred set of values of the critical indices.

\section{Conventional LPA versus pseudo-LPA\label{VersusPseudoLPA}}

In this section, we first show that the argument of Hasenfratz-Hasenfratz 
\cite{2085}, by which the RG flow projected on (\ref{2}) would imply $\eta
=0 $ if $\bar{z}$ is kept unaltered by the flow of $U$, is artificial and
reduces to a triviality that poorly \textquotedblleft
justifies\textquotedblright\ the default argument. Then we briefly
illustrate that, correctly treated, the projection of the ERGE on (\ref{2})
gets an \textquotedblleft intermediate order\textquotedblright\ [between the 
$O\left( \partial ^{0}\right) $ and $O\left( \partial ^{2}\right) $] of the
derivative expansion that we name pseudo-LPA. This partial $O\left( \partial
^{2}\right) $-order differs from the approximation introduced in \cite{3642}
in that we try to account for the reparametrization invariance.

\subsection{Invalidity of the conventional argument\label{Invalidity}}

To get the RG flow equations of the Wilson-Polchinski ERGE correctly
projected on (\ref{2}), suffices to consider the complete $O\left( \partial
^{2}\right) $-order equations available in the literature as, e.g., in \cite%
{3491,3836}, and to impose within them that the kinetic term is a pure
number that remains constant along a RG flow of the potential.

For example, let us consider Eqs. (12) of \cite{3836} for the derivative $%
f\left( \phi ,t\right) =U^{\prime }\left( \phi ,t\right) $ and a function $%
Z\left( \phi ,t\right) $ reduced to $\bar{z}\left( t\right) $, it comes: 
\begin{eqnarray}
\dot{f} &=&f^{\prime \prime }-2\,f\,f^{\prime }-\frac{D-2+\eta \left(
t\right) }{2}\,\,\phi \,f^{\prime }+\frac{D+2-\eta \left( t\right) }{2}\,f
\label{PLPAV} \\
\overset{\cdot }{\bar{z}}\left( t\right) &=&-\eta \left( t\right) \,\,\bar{z}%
\left( t\right) +2\,B\,f^{\prime }\left( 0,t\right) ^{2}-4\,\left[ \bar{z}%
\left( t\right) -1\right] \,f^{\prime }\left( 0,t\right)  \label{PLPAZ}
\end{eqnarray}%
where $B$ is a constant parameter depending on the choice of cutoff function.

Up to inessential changes, Eq. (\ref{PLPAV}) is the same flow equation as (%
\ref{LPA}) of the pure potential $U$ discussed previously in section \ref%
{RevisedLPA}. Eq. (\ref{PLPAZ}) shows that the flow of $U$ induces a flow of 
$\bar{z}$ so that keeping it constant, i.e. imposing $\overset{\cdot }{\bar{z%
}}\left( t\right) =0$, yields:%
\begin{equation}
\eta \left( t\right) \,\,\bar{z}\left( t\right) -2\,B\,f^{\prime }\left(
0,t\right) ^{2}+4\,\left[ \bar{z}\left( t\right) -1\right] \,f^{\prime
}\left( 0,t\right) =0  \label{P1}
\end{equation}%
this condition considered at a fixed point of (\ref{PLPAV}) may be rewritten
as:%
\begin{equation}
\eta ^{\ast }\,\,\bar{z}-2\,B\,\gamma ^{2}+4\,\left[ \bar{z}-1\right]
\,\gamma =0  \label{P0}
\end{equation}%
where, as seen in section \ref{RevisedLPA} (footnote \ref{gamma}), $\gamma
=f^{\ast \prime }\left( 0\right) =U^{\ast \prime \prime }\left( 0\right) $
is a function of $\eta ^{\ast }$ as that given implicitly by the lines of
fixed points $\mu _{n}^{\ast }$ drawn in fig. (\ref{figPF1}) where $U^{\ast
\prime \prime }\left( 0\right) $ was playing the role of $\bar{z}$. For $%
\eta ^{\ast }=0$ and $\bar{z}=1$ (the conventional values), Eq. (\ref{P0})
implies $U^{\ast \prime \prime }\left( 0\right) =0$. Fig. (\ref{figPF1})
shows that this is not possible along the line of W\thinspace F fixed points 
$\mu _{1}^{\ast }$ except trivially at $D=4$ where the fixed point is
Gaussian$.$

The conventional argument of \cite{2085} actually relies upon the arbitrary
requirement that no contribution of $U$ must alter the flow of $\bar{z}$ so
that the right-hand-side of (\ref{PLPAZ}) would be reduced to the first term
exclusively, thus implying $\eta \left( t\right) \equiv 0$ for a constant $%
\bar{z}$. Clearly, this is only an illustration of an obvious fact: the
non-necessity of renormalizing the field (here forced by the obligatory
absence of contribution coming from $U$), induces $\eta \left( t\right)
\equiv 0$ (what is true by definition). We add that, not only truncation (%
\ref{2}) is incompatible with a pure $O\left( \partial ^{0}\right) $-order,
it is also in contradiction with the default argument --by which $\eta
\left( t\right) $ is absent because there is \textsl{no momenta} in the LPA.
Actually, the conventional argument is merely artificial. As shown in
section \ref{RevisedLPA}, it is the basis of a conventional LPA which is
misleading concerning the concept of reparametrization invariance and in
contradiction with the logic of the derivative expansion.

\subsection{Pseudo-LPA\label{PseudoLPA}}

Considered as an actual truncation of the action $S\left[ \phi \right] $,
Eq. (\ref{2}) gives access to an intermediate approximate order of the ERGE
[between $O\left( \partial ^{0}\right) $ and $O\left( \partial ^{2}\right) $%
]. By allowing the coefficient of the kinetic term to flow (whereas it
remains independent of $\phi $) one obtains the partial truncation first
used by Tetradis and Wetterich \cite{3642} in order to easily have $\eta
^{\ast }\neq 0$ in an \textquotedblleft improved\textquotedblright\
(conventional) LPA. But the way of determining $\eta ^{\ast }$, in the
original proposal, is limited to the ERGE for the effective average action $%
\Gamma \left[ \varphi \right] $ (for a review see \cite{4700}). This is
because one utilizes the available momentum dependence of the exact
propagator $\Gamma ^{(2)}$ to determine the function $\bar{z}\left( t\right) 
$ yielding a value for $\eta ^{\ast }$. That way of doing is not convenient
to the Polchinski ERGE, with which an easy access to $\Gamma ^{(2)}$ is not
possible. Moreover, the Tetradis and Wetterich approach does not give a
clear account of the reparametrization invariance (being in the spirit of
the conventional LPA criticized in the present paper).

Let us look at truncation (\ref{2}) for the Polchinski ERGE by strictly
applying the basic rules of the RG theory.

The RG rules prescribe a field renormalization in order to maintain constant
one term of the action. With truncation (\ref{2}), we choose it to be the
kinetic term\footnote{%
With a complete $O\left( \partial ^{2}\right) $-order we could have pursued
the process of maintaining constant the quadratic term of the action
(instead of the kinetic term). In the present pseudo-LPA this procedure
would give nothing new compared to the pure LPA. This underlines the
particular character of that truncation.\label{pseudoFoot}} (the only
momentum-dependent monomial of the approximation). Hence we get Eq. (\ref{P1}%
) and, at a fixed point, Eq. (\ref{P0}). Considering $\bar{z}$ as a free
parameter at hand, $\eta ^{\ast }$ and $\gamma $ appear to be functions of $%
\bar{z}$. We thus get lines of fixed points parametrized by $\bar{z}$. The
relation between $\eta ^{\ast }$ and $\gamma $ being unchanged compared to
the LPA [Eq. (\ref{PLPAV}) is the derivative with respect to $\phi $ of (\ref%
{LPA0})], we have:%
\begin{equation*}
\gamma \left( \bar{z}\right) =\gamma _{\mathrm{LPA}}\left[ \eta ^{\ast
}\left( \bar{z}\right) \right]
\end{equation*}%
where $\gamma _{\mathrm{LPA}}\left[ \eta ^{\ast }\right] $ has been
determined at leading order in section \ref{FixedPoints} [implicitly through
the lines of fixed points displayed by fig. (\ref{figPF1})].

It is then easy to seek for a vestige of the reparametrization invariance
eventually displayed by the new lines of fixed points (parametrized by $\bar{%
z}$). Suffices to look at possible values of $\eta ^{\ast }$ where $\mathrm{d%
}\eta ^{\ast }/\mathrm{d}\bar{z}=0\,$.

Differentiating (\ref{P0}) with respect to $\bar{z}$, we get:%
\begin{equation*}
\frac{\mathrm{d}\eta ^{\ast }}{\mathrm{d}\bar{z}}\,\left\{ \,\bar{z}%
-4\,B\,\gamma _{\mathrm{LPA}}\gamma _{\mathrm{LPA}}^{\,\,\prime }+4\,\left[ 
\bar{z}-1\right] \,\gamma \,_{\mathrm{LPA}}^{\prime }\right\} +\eta ^{\ast
}+4\,\gamma _{\mathrm{LPA}}=0
\end{equation*}%
where $\gamma _{\mathrm{LPA}}^{\,\prime }=\mathrm{d}\,\gamma _{\mathrm{LPA}}/%
\mathrm{d}\eta ^{\ast }$. Finally, imposing the required condition gives a
preferred value of $\eta ^{\ast }$ defined by:%
\begin{equation*}
\eta _{\mathrm{opt}}^{\ast }=-4\,\gamma _{\mathrm{LPA}}\left( \eta _{\mathrm{%
opt}}^{\ast }\right)
\end{equation*}

Notice that this condition is independent of the choice of the cutoff
function, contrary to what is observed with the complete $O\left( \partial
^{2}\right) $-order \cite{3491,3836} (for another difference with the
complete order, see footnote \ref{pseudoFoot}).

In terms of the quantity $\mu $ defined by (\ref{mu}) and taking into
account the change of field variable (\ref{ChangePhi}), this condition
writes, for $D=3$ and $\gamma _{\mathrm{LPA}}=D\,z$ (see footnote \ref{gamma}%
)%
\begin{equation*}
\mu ^{\ast }=\frac{1-12\,z}{6}
\end{equation*}

From the calculations done in section \ref{FixedPoints}, we obtain:%
\begin{eqnarray*}
\eta _{\mathrm{opt}}^{\ast } &\simeq &0.269 \\
\nu &\simeq &0.649
\end{eqnarray*}%
which are not excellent values. This result may however be considered as an
improvement compared to the LPA for which no vestige of the
reparametrization invariance was observed.

\section{Summary and conclusion\label{Conclusion}}

We have justified the presence of a non-vanishing value of $\eta $ in the
LPA as a strict consequence of the general principles of the RG theory.
Without field renormalization, as usually prescribed in the conventional LPA
(with $\eta =0$), the approximation could not be considered as the genuine $%
O\left( \partial ^{0}\right) $-order of the derivative expansion. If no
particular estimate of $\eta $ can actually be proposed at this order, $\eta 
$ is not an arbitrary parameter, it varies monotonously (within some limits)
on changing the normalization of the field as a consequence of a perfect
breaking of the reparametrization invariance. The situation is coherent with
the idea that, if the derivative expansion converges then, at least, a
progressive and smooth restoration of the invariance must be observed from
the few first terms of the expansion (this was not possible with the
conventional view). We have done explicit calculations of the lines of
fixed points generated by the change of the normalization of the field by a
constant $z$ using both purely numerical and quasi-analytic methods in order
to offer the possibility to anyone to easily redo the calculations. We also
emphasize that, despites the minimum observed (for $D=3$) in the evolution
of the critical exponent $\nu $ on varying $z$, the principle of minimal
sensitivity cannot be applied being not compatible with a possible vestige
of the reparametrization invariance in the LPA. We have shown (in section %
\ref{Invalidity}) that it is purely artificial the conventional argument
stating that $\eta $ should vanish if one keeps the kinetic term unchanged
along a RG flow of the potential. We have illustrated (in appendix \ref%
{Movable}) the respective roles of the movable and fixed singularities of
the fixed point solutions in the convergence property of the quasi-analytic
methods of integration utilized in the study.

\appendix{}

\section{Movable singularity and convergence of the simplistic method \label%
{Movable}}

In this appendix we illustrate the role of the movable singularity of the
solution of the fixed point equation of the LPA in the convergence and
efficiency of the simplistic method (see section \ref{TSM}).

Let us consider the fixed point equations of respectively the RG flow (\ref%
{LPA0}) of $U$ and the RG flow (\ref{LPAMorris0}) of the Legendre
transformed potential $V$ [see (\ref{TL1}-\ref{TL3})]. For $D=3$ and $\eta
\left( t\right) =0$ the common value of the connection parameter $k^{\ast
}=U^{\ast }\left( 0\right) =V^{\ast }\left( 0\right) $, corresponding to the
respective regular fixed point solutions, is known with a huge number of
digits \cite{6319} to be:%
\begin{equation}
k^{\ast }=0.07619940081234\cdots  \label{kstar}
\end{equation}

That value corresponds precisely to the only solution of the two-point
boundary problem (\ref{FP}-\ref{cond2}) with $\mu ^{\ast }=1/6$. If one
forgets about the condition at infinity, then it exists a solution involving
a singularity for each value\footnote{%
That singularity having a location which depends on the value of $k$, is
named movable singularity.} of $k=U\left( 0\right) $ different from $k^{\ast
}$. Hence getting the value (\ref{kstar}) may be viewed as the consequence
of pushing that movable singularity to infinity.

In addition to a movable singularity, a solution of (\ref{FP}-\ref{cond1})
displays also fixed singularities\footnote{%
The locations of which vary very slowly with $k$.}. Potentially, those
singularities control the convergence properties of the Maclaurin series of
the ultimate $U^{\ast }$.

We have performed Pad\'{e} approximants on the Maclaurin series of solutions
for various $k$ of the fixed point equations in the two cases of $U$ and $V$%
. The complex zeros of the denominators of the corresponding rational
fractions give an approximate image of the location of the singularities in
the complex plane of the variables \underline{$x$}$=\phi ^{2}$ and $%
x=\varphi ^{2}$ of respectively $U$ and $V$ .

\begin{figure}[tbp]
\begin{center}
\includegraphics*[width=10cm]{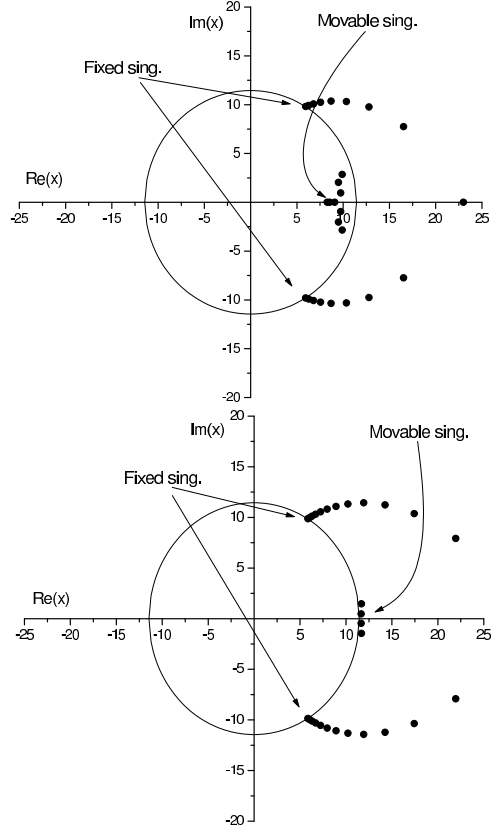}
\end{center}
\caption{Evolution of the disposition, in the complex plane of the variable $%
x=\protect\varphi ^{2}$, of the singularities of the solutions $V^*\left( 
\protect\varphi \right)$ of the fixed point equation, for $\protect\eta^*=0$ and $D=3$%
, of the RG flow Eq. (\protect\ref{LPAMorris0}) in the process of
determining $k^{\ast }=$ $V^{\ast }\left( 0\right) $ the value of which is
given by Eq. (\protect\ref{kstar}). Two steps are shown: $k=k^{\ast
}+8\times 10^{-4}$ (top) and $k=k^{\ast }-8\times 10^{-12}$ (bottom). As one
approaches $k^{\ast }$, the movable singularity is pushed on the right
(ideally up to infinity). The efficiency of the simplistic method ceases
when the movable singularity is about to leave the disc of convergence of
the Maclaurin series (determined by the location of the fixed singularity
the closest to the origin). The method yields a rather high accuracy on the
estimate of $k^*$ with 9 accurate figures. The radius of convergence of the
Maclaurin series for $k=k^{\ast }$ is $R=11.449$.}
\label{FigSinLitim}
\end{figure}

Figure (\ref{FigSinLitim}) shows the singularity structure of $V$ for two
values of $k$ on approaching $k^{\ast }$. One clearly sees that the movable
singularity still lies within the circle of convergence of the Maclaurin
series though $k$ is already close to $k^{\ast }$. On approaching closer to $%
k^{\ast }$ the movable singularity is pushed to the right and the simplistic
method ceases to converge when the singularity comes out of the disc of
convergence of the series. Concretely the method provides an estimate of $%
k^{\ast }$ with 9 accurate figures.

\begin{figure}[tbp]
\begin{center}
\includegraphics*[width=10cm]{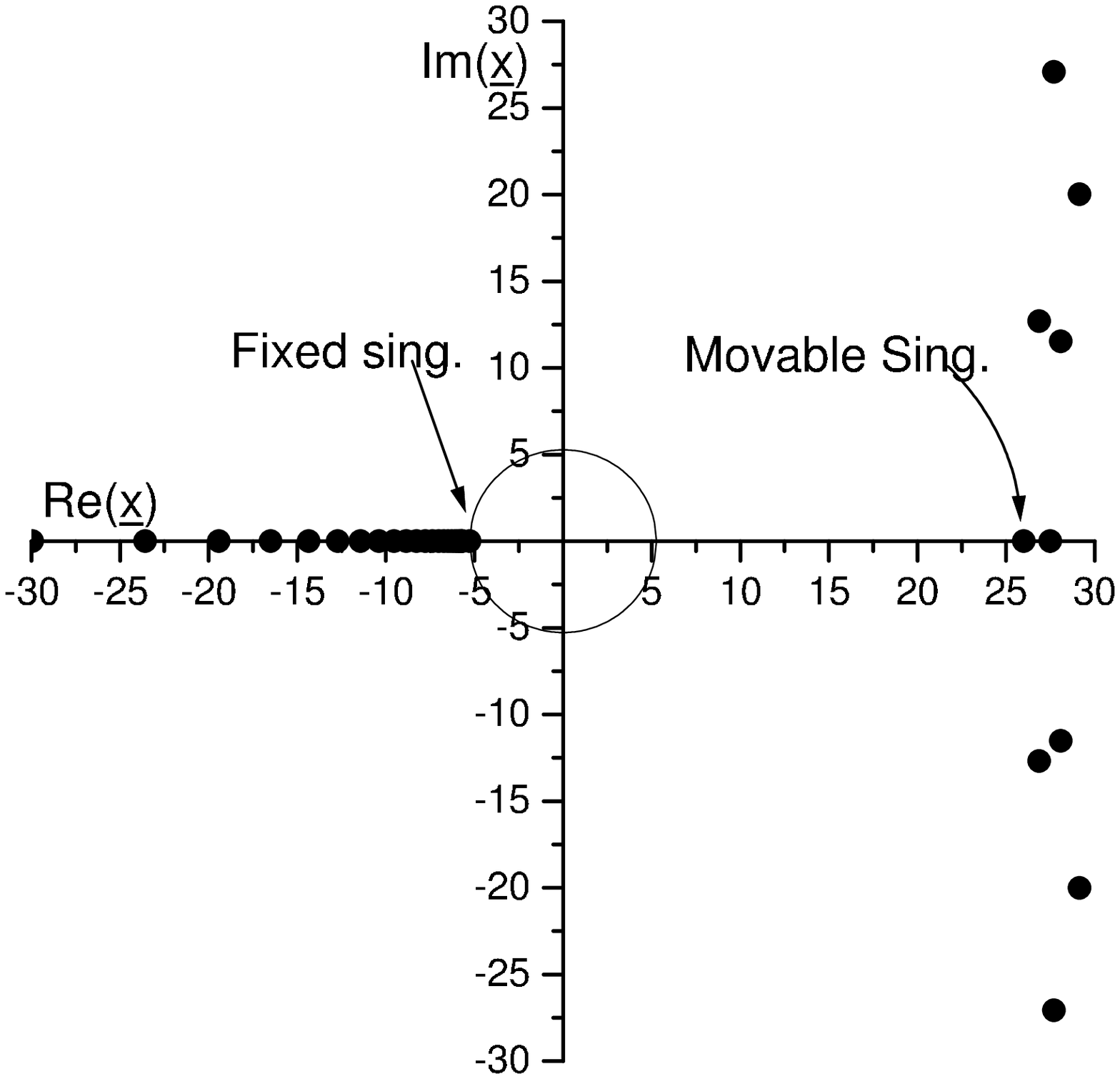}
\end{center}
\caption{Disposition, in the complex plane of the variable \protect%
\underline{$x$}$=\protect\phi ^{2}$, of the singularities of the solution of
Eq. (\protect\ref{FP}), for $\protect\eta^*=0$ and $D=3$, in the process of
determining $k^{\ast }=$ $U^{\ast }\left( 0\right) $ given by Eq. (\protect
\ref{kstar}). Same presentation as in fig. (\protect\ref{FigSinLitim}). The
step corresponds to $k=k^{\ast }+4\times 10^{-3}$ which is relatively far
from $k^{\ast }$ whereas the movable singularity is already outside the disc
of convergence of the Maclaurin series. The simplistic method does not work
in that case. The radius of convergence of the Maclaurin series for $%
k=k^{\ast }$ is $R=5.2719.$}
\label{figSingPol}
\end{figure}

On the contrary, fig. (\ref{figSingPol}) shows that the movable singularity
is already well outside the circle of convergence of the Maclaurin series of 
$U$ when $k$ is still far from $k^{\ast }$. In that case the simplistic
method cannot even give a poor estimate of $k^{\ast }$.

\end{document}